\documentclass[aps,prc,11pt,tightenlines,
showpacs,preprintnumbers,amsmath,amssymb,
superscriptaddress,
a4paper,nofootinbib,titlepage,nolongbibliography]{revtex4-1}

\usepackage{nicefrac}

\usepackage{fouriernc}

\usepackage{natbib}
\usepackage{amssymb,amsbsy,amsmath,amsfonts}
\usepackage{graphicx}
\usepackage{epsf,epsfig,float,latexsym,amsthm,fancyhdr,rotating}
\usepackage{graphics,psfrag,longtable}
\usepackage{slashed}
\usepackage{comment}
\usepackage[colorlinks,citecolor=blue,linktoc=all,linkcolor=cyan,urlcolor=blue]{hyperref}
\usepackage{upgreek}

 \usepackage{setspace}

\usepackage[normalem]{ulem}

\def\beq{\begin{equation}}
\def\eeq{\end{equation}}
\def\bea{\begin{eqnarray}}
\def\eea{\end{eqnarray}}
\def\beqa{\begin{equation}\begin{array}{l}}
\def\eeqa{\end{array}\end{equation}}

\def\Figref#1{Fig.~\ref{fig:#1}}

\def\barr{\left(\begin{array}{c}}
\def\earr{\end{array}\right)}
\def\bmat{\left(\begin{array}{cc}}
\def\emat{\end{array}\right)}

\def\3d{3-D}

\begin{document}
\title{Muonic-Atom Spectroscopy and Impact on  Nuclear Structure and Precision QED Theory}

\author{Aldo~Antognini}
\email{aldo.antognini@psi.ch}
\affiliation{Institute for Particle Physics, ETH, 8093 Zurich, Switzerland
}
\affiliation{Laboratory for Particle Physics, Paul Scherrer Institute, 5232 Villigen-PSI, Switzerland 
}

\author{Sonia~Bacca}
\affiliation{Institute  of  Nuclear  Physics,
Johannes Gutenberg-Universit\"at Mainz, 55099 Mainz, Germany
}
\affiliation{PRISMA+ Cluster  of  Excellence, Johannes  Gutenberg-Universit\"at  Mainz, 55099 Mainz, Germany}

\author{Andreas~Fleischmann}
\affiliation{Universität Heidelberg, Kirchhoff‐Institut für Physik, INF 227, 69120 Heidelberg
}

\author{Loredana~Gastaldo}
\affiliation{Universität Heidelberg, Kirchhoff‐Institut für Physik, INF 227, 69120 Heidelberg
}

\author{Franziska~Hagelstein} 
\email{hagelste@uni-mainz.de}
\affiliation{Laboratory for Particle Physics, Paul Scherrer Institute, 5232 Villigen-PSI, Switzerland 
}
\affiliation{Institute  of  Nuclear  Physics,
Johannes Gutenberg-Universit\"at Mainz, 55099 Mainz, Germany
}
\affiliation{PRISMA+ Cluster  of  Excellence, Johannes  Gutenberg-Universit\"at  Mainz, 55099 Mainz, Germany}

\author{Paul~Indelicato}
\affiliation{Laboratoire Kastler Brossel, Sorbonne Universit\'{e}, CNRS, ENS-PSL Research University, Coll\`{e}ge de France, Case 74, 4, place Jussieu, 75005 Paris, France}

\author{Andreas~Knecht}
\affiliation{Laboratory for Particle Physics, Paul Scherrer Institute, 5232 Villigen-PSI, Switzerland 
}

\author{Vadim~Lensky} 
\affiliation{Institute  of  Nuclear  Physics,
Johannes Gutenberg-Universit\"at Mainz, 55099 Mainz, Germany
}

\author{Ben~Ohayon}
\email{bohayon@ethz.ch}
\affiliation{Institute for Particle Physics, ETH, 8093 Zurich, Switzerland
}
\affiliation{Physics Department, Technion—Israel Institute of Technology, Haifa 3200003, Israel
}

\author{Vladimir Pascalutsa} 
\affiliation{Institute  of  Nuclear  Physics,
Johannes Gutenberg-Universit\"at Mainz, 55099 Mainz, Germany
}

\author{Nancy~Paul}
\affiliation{Laboratoire Kastler Brossel, Sorbonne Universit\'{e}, CNRS, ENS-PSL Research University, Coll\`{e}ge de France, Case 74, 4, place Jussieu, 75005 Paris, France}

\author{Randolf~Pohl}
\affiliation{Institut für Physik, QUANTUM, Johannes Gutenberg-Universität Mainz, 55099 Mainz,
Germany}
\affiliation{PRISMA+ Cluster  of  Excellence, Johannes  Gutenberg-Universit\"at  Mainz, 55099 Mainz, Germany}

\author{Frederik~Wauters}
\affiliation{Institute  of  Nuclear  Physics,
Johannes Gutenberg-Universit\"at Mainz, 55099 Mainz, Germany
}
\affiliation{PRISMA+ Cluster  of  Excellence, Johannes  Gutenberg-Universit\"at  Mainz, 55099 Mainz, Germany}

\begin{abstract}
Recent progress in laser and x-ray spectroscopy of muonic atoms offers promising long-term possibilities at the intersection of atomic, nuclear and particle physics. 
In muonic hydrogen, laser spectroscopy measurements will determine the ground-state hyperfine splitting (HFS) and additionally improve the Lamb shift by a factor of 5. Precision spectroscopy with cryogenic microcalorimeters has the potential to significantly improve the charge radii of the light nuclei in the $Z=3-8$ range. Complementary progress in precision should be achieved on the theory of nucleon- and nuclear-structure effects. The impact of this muonic-atom spectroscopy program will be amplified by the upcoming results from H and He$^+$ spectroscopy, simple molecules such as HD$^+$ and Penning trap measurements. In this broader context, one can test ab-initio nuclear theories, bound-state QED for two- or three-body systems, and determine fundamental constants, such as the Rydberg ($R_\infty$)  and the fine- structure ($\alpha$) constants.
\end{abstract}
\maketitle


\section{Introduction} 
\begin{figure*}[h]
\includegraphics[trim=55 35 70 58,clip,width=0.85\textwidth]{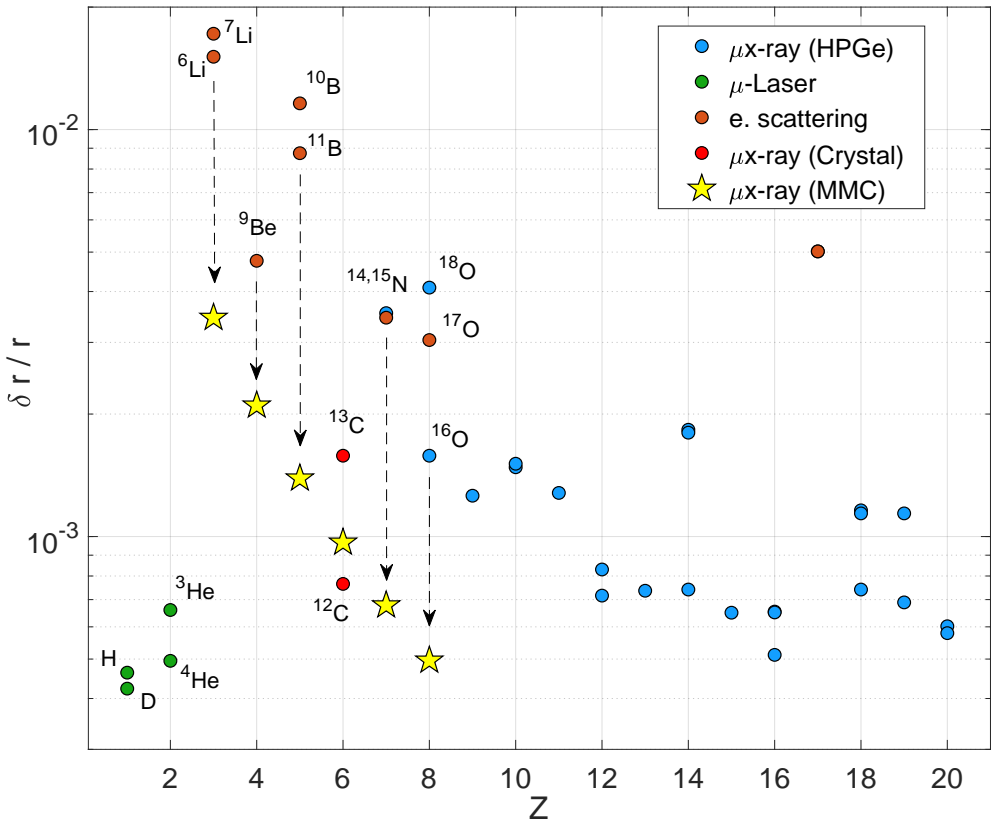}
\caption{
Relative uncertainty in the most precise RMS charge radii of the light stable nuclei. The stars denote the achievable precision when $E_{2P-1S}$ transitions in muonic atoms are measured with 10\,ppm accuracy.}
\label{fig:rad}
\end{figure*}

Muonic atoms --- hydrogen-like atoms with the electrons replaced by a muon --- 
have an enhanced sensitivity to the structure of the atomic nucleus.
For light muonic atoms, the enhancement factor, as compared to regular atoms, is of the order of $(m_\mu/m_e)^3\approx 10^7$, making them a prime laboratory for studies of nuclear structure and in particular for the determination of RMS nuclear charge radii (see Fig.~\ref{fig:rad}).

The proton charge radius $r_p$ was measured via laser spectroscopy of the 2S-2P transition in muonic hydrogen ($\mu$H) by the CREMA Collaboration \cite{Pohl:2010zza,Antognini:1900ns}, appearing to be more than $5\sigma$ away from the value based on $ep$ scattering and H spectroscopy. 
This \textit{proton-radius puzzle} triggered a wealth of activity at the intersection of nuclear, particle, and atomic physics, reaching out to  physics beyond the Standard Model (see Refs.~\cite{Carlson:2015jba,Karr:2020wgh,Gao:2021sml,Peset:2021iul,Antognini:2022xoo}, for recent reviews). 
Today, more than a decade after this puzzle began, new determinations of the proton radius are available that corroborate the results from measurements with muonic atoms. 
However, there are still some tensions that call for further experimental determinations~\cite{Brandt:2021yor}, and a new round of experiments is underway \cite{Strauch:2018ros,COMPASSAMBERworkinggroup:2019amp,PRad:2020oor}.

Laser spectroscopy of muonic atoms has also determined the radii of the deuteron and $^{3,4}$He with a very high accuracy $<10^{-3}$~\cite{Pohl1:2016xoo,2017-3He,Krauth:2021foz}.
Extending this method to heavier elements is highly challenging. Nevertheless, it is considered in Li and Be by the CREMA collaboration \cite{Schmidt:2018kjc}.
For $Z>8$, most radii of stable isotopes have been determined by measuring the energy of x-rays emitted in the cascade of muons to the ground level~\cite{2004-Fricke}. However, for nuclei in the range $Z=3-8$, with respective 2p-1s transition energies of $20-150\,$ keV, the energy resolution of solid state detectors limits the precision of such a measurement.
In light of this, most radii of stable isotopes in this range are determined via elastic electron scattering, and are also the least well-known in the nuclear chart, with accuracies spanning $0.3-1.4\%$ (see Fig.~\ref{fig:rad}).

A precise knowledge of charge radii is important for benchmarking \textit{ab initio} nuclear theory. When combined with measurements in electronic atoms, they enable tests of bound-state QED (BSQED), determining fundamental constants and searching for new physics (see \Figref{Flow-chart}).
Such tantalizing prospects rely on new experiments with muonic atoms as well as developments in atomic and nuclear theory as discussed below.

\section{Benchmarking nuclear theory} 

Precise knowledge of light nuclear masses and radii is necessary to set rigorous benchmarks for nuclear structure theory. Although masses are measured with high precision using Penning traps, they are not an effective discriminator between theories \cite{2022-Light}. Radii reveal more about the underlying theory, but are much harder to calculate due to their sensitivity to the long range behaviour of the nuclear wave functions \cite{2019-Calc}.

Recent progress for nucleon-nucleon currents and form factors of light nuclei has been reported in Refs.~\cite{Filin:2019eoe,Filin:2020tcs}. Combining theory predictions of the deuteron and alpha particle structure radii  from chiral effective field theory and their  charge radii from spectroscopy, one obtains precise predictions of the proton and neutron charge radii. A prediction of the isoscalar combination of the triton and helion structure radii, together with the helion charge radius from $\mu^3$He$^+$, will also predict the charge radius of the triton, allowing one, in turn, to test nuclear structure calculations.
Going to higher Z, \textit{ab-initio} calculations of charge radii using nuclear interactions derived from chiral effective field theory have advanced enough to reach a high accuracy in the region $A=3$ to $A=18$ \cite{2015-QMC,2020-CEFT}, which could then be systematically improved based on comparison with experiments. 

Another benchmark comes from the evolution of neutron skins with isospin, which remains a large open question \cite{2021-Skin}. As neutron skins are highly difficult to measure, the difference in charge radii of mirror nuclei could be used as a surrogate \cite{2018-Surr}. Being a differential measurement, high accuracy on the individual absolute radii ($<0.1$\%) is needed. Extending well-measured radii of mirror pairs to lighter nuclei: $^3$H-$^3$He, $^7$Li-$^7$Be and $^8$Li-$^8$B, will test nuclear theory at a large relative neutron excess. It would also provide valuable input into determining the parameters of the equation of state of nuclear matter \cite{2017-EOS}. 
To do this, one must combine the absolute radii of one isotope of He, Li, Be, and B, with optical measurements of isotope shifts.
Such measurements were accomplished for $^{8}$Li \cite{2011-LiIS} and $^{7}$Be \cite{2009-Be}, while ongoing efforts are focused on $^3$H \cite{Schmidt:2018kjc} and ${^8}$B \cite{2017-B8}. 

\section{Precision Tests of Bound-State QED}

\begin{figure}[t]
\includegraphics[width=0.9\textwidth]{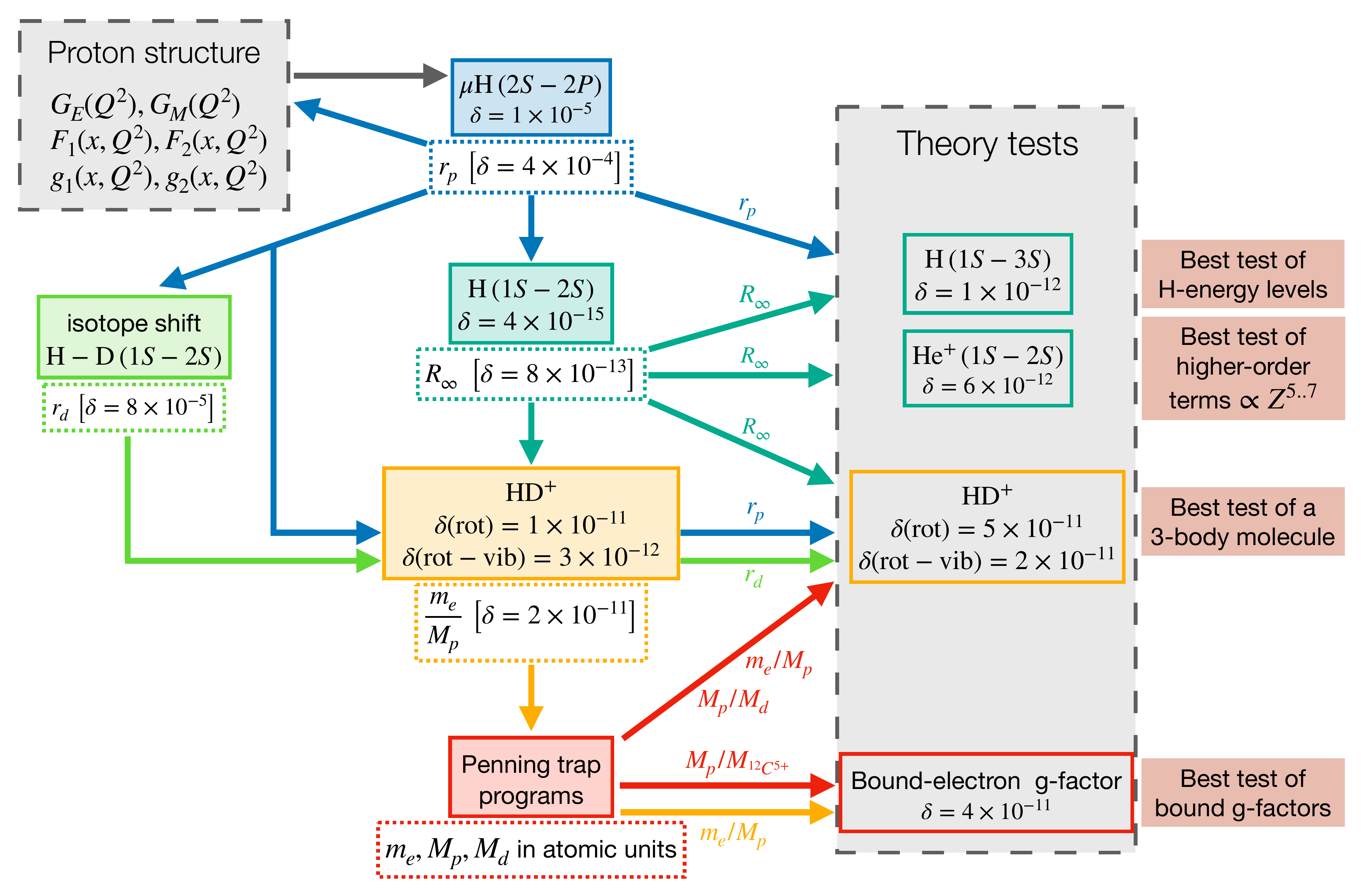}
\caption{Interplay of experiments to improve fundamental constants and bound-state QED tests.}
        \label{fig:Flow-chart}
\end{figure}

The simplicity of two- and three-body atomic/molecular systems combined with the precision of laser spectroscopy permits unique confrontations between theory and experiments, see \Figref{Flow-chart}.
The predictive power of BSQED which describes these systems, however, depends on the knowledge of fundamental constants such as the masses of the involved particles, the Rydberg constant $R_\infty$, and the nuclear charge radii.
The precise nuclear charge radii obtained from $\mu$H, $\mu$D~\cite{Pohl1:2016xoo} and $\mu$He$^+$~\cite{Krauth:2021foz} are therefore not only benchmarks to understand the hadron and nuclear structure but also allow to push the confrontation between theory and experiments in two- and three-body atomic/molecular systems.

Comparing the  1S-2S transition  measurement in H~\cite{Parthey:2011lfa} with its theory prediction that makes use of the precise $r_p$ value from $\mu$H, leads to a precise determination of $R_\infty$  with a relative accuracy of  $8\times10^{-13}$.
By considering in addition to the 1S-2S transition, more measurements in H allows to confront the theory in hydrogen (with all fundamental constant fixed) to measurements~\cite{Brandt:2021yor, Grinin:2020:Science_H1S3S}. In this way the hydrogen theory  has been verified at the $1\times 10^{-12}$ level.
Similarly, the alpha particle charge radius from $\mu$He$^+$ spectroscopy~\cite{Krauth:2021foz} will be needed when the ongoing efforts to measure the 1S-2S in He$^+$ will be completed~\cite{Krauth:2019kei, PhysRevA.79.052505}.
With $R_\infty$ obtained by combining $\mu$H and H measurements, the confrontation theory-experiment in He$^+$ provides a bound-state QED test particularly sensitive to challenging higher-order corrections scaling with $Z^{5..7}$~\cite{Yerokhin2018, Karshenboim:2019iuq}.

Calculations of the energy levels in three-body systems such as helium and helium-like ions are rapidly advancing to the point where nuclear-size effects become important \cite{2010-HLI,2017-He,2022-HLI}.
With these developments, the ongoing experiments in helium-like ions spanning Li-II to C-V \cite{2020-HLI_Li, 2020-BHLI, 2020-COALA} would be sensitive to charge radii at the level of $10^{-3}$.
As with the lighter systems, the efforts with electronic atoms would greatly benefit from new measurements with muonic atoms.
Such combination would allow to test BSQED for three-body systems at higher Z. Conversely, one could benchmark nuclear polarizability calculations in systems heavier than He. At the level in which these calculations are reliable, one could search for new physics affecting the muonic and/or electronic observables~\cite{2017-YS,2018-X,2022-PESET}.

Other interesting three-body systems are molecular ions. Precision values of the proton and deuteron radii play an increasingly important role also for the precise predictions of the HD$^+$ energy levels given the spectacular advances both in theory and experiment~\cite{Karr:2020jbg, Alighanbari2020, Patra2020, Kortunov:2021rfe}. With $R_\infty$ and nuclear charge radii obtained by combining $\mu$p and H measurements,  and the electron/proton and proton/deuteron mass ratios from Penning trap measurements,  the HD$^+$ theory has been verified  at the    $2\times 10^{-11}$ level, representing the best test of a three-body system.
Conversely, the comparison of theory with experiment in HD$^+$ can be used to improve on the  electron  mass (electron/proton mass ratio) while using charge radii and $R_\infty$ as derived from $\mu$H and H.
Combined with other Penning trap measurements~\cite{Sturm:2014bla},  the new value of the electron mass allows testing bound electron g-factors~\cite{Zatorski:2017vro}  on the $4\times 10^{-11}$ level of accuracy. In the future, it will also impact the determination of $\alpha$.
Hence, the HD$^+$ spectroscopy links precision laser spectroscopy of hydrogen-like atoms and light muonic atoms with the rich Penning trap program that includes testing bound electron g-factors, determination of fundamental constants and precision mass spectrometry. 

Spectroscopy of heavier muonic atoms can also be used to test strong-field BSQED by focusing on transitions between Rydberg states where there is little to no overlap between the muon and nuclear wavefunctions \cite{2021-Paul}.
The relevant energies are in the $3-200\,$keV range, suitable for high-resolution detection via cryogenic microcalorimeters.
The dominant QED effect in these systems is the vacuum polarization, providing a unique window into this effect, as well as new features of muonic atom dynamics \cite{2021-Okumura}. Future BSQED studies with muonic atoms will also be highly complementary to planned programs for high-resolution antiprotonic atom spectroscopy at ELENA for atomic and nuclear structure studies.  

\section{Theoretical developments: Nucleon and Nuclear-Structure Effects} 

The limiting uncertainty in calculations of the energy levels of muonic atoms comes from nuclear- and nucleon-structure effects.
Improved theory predictions, especially for the polarizability contributions, are needed  to match up with the present experimental precision for $\mu$D and $\mu$He$^+$, as well as to compete with the upcoming experiments in $\mu$H, discussed in Section \ref{Sec:MuonicExp}. 

For $\mu$H, the proton finite-size and polarizability effects can be either calculated in $\chi$EFT, or in a  dispersive framework from data on the proton form factors and structure functions.  So far, these two frameworks agreed well in the proton-polarizability contribution to the Lamb shift, but much less so in  the hyperfine splitting (hfs) \cite{Pascalutsa:2022ckb}. A re-evaluation with new data from the Jefferson Lab ``Spin Physics Program'' \cite{Chen:2008ng}, e.g., from the g2p experiment \cite{JeffersonLabHallAg2p:2022qap}, is underway. Improved  $\chi$EFT calculations would be helpful,  as well the proposed lattice QCD calculations \cite{Hagelstein:2020awq,CSSMQCDSFUKQCD:2022fzy,Fu:2022fgh}. 

For $\mu$D and $\mu$He$^+$, both the nucleon- and nuclear-structure effects enter. They are calculated in the data-driven dispersive framework \cite{Carlson:2013xea,Carlson:2016cii}, and in a number of theoretical approaches, such as pionless effective field theory \cite{Lensky:2022bid}, 
chiral effective field theory~\cite{HERNANDEZ:2018} or phenomenological Hamiltonians~\cite{Ji:2018ozm}. For the $\mu$D Lamb shift, recent theory predictions are in agreement, and through the H-D isotope shift are consistent with the proton radius from $\mu$H,  see, e.g., Ref.~\cite{Lensky:2022bid}.

In $\mu$He$^+$ the nucleon contributions become increasingly important.
Consider for instance the two-photon-exchange (2PE) contribution to the $\mu^4$He$^+$ Lamb shift: $E^{\langle \mathrm{2PE} \rangle A+N}_{2P-2S}=9.34(20)_\mathrm{N}(11)_\mathrm{A}\,\mathrm{meV}$ \cite{Diepold:2016cxv}, with $A$ and $N$ the nuclear and nucleon contributions.  
The nucleon part  gives a sizeable contribution to the uncertainty. In order to reduce the uncertainty, one has to improve on the following quantities: the nucleon-polarizability contribution (primarily the neutron), the nuclear-polarizability contribution  
(presently limited by the spread from various implementations of the few-nucleon dynamics~\cite{NEVODINUR:2016}), and the electric form factor needed to compute the elastic part~\cite{Diepold:2016cxv, Ji:2018ozm}. More data on the neutron polarizabilities could be obtained at MAMI (Mainz) and HIGS (Duke).

Given the prospects to study $\mu$Li, we expect the nuclear polarizability contributions to become extremely important. The latter can be and will be tackled within \textit{ab-initio} calculations~\cite{LiMuli:2020}.
For the heavier nuclei, similarly to what was implemented in $\mu$C~\cite{1982-12C}, one could use input from total integrated nuclear photoabsorption cross-sections to estimate the polarizability.

\section{New experiments with muonic atoms}\label{Sec:MuonicExp}

There are three collaborations aiming at the measurement of the ground-state HFS in $\upmu$H: one collaboration at J-PARC (Japan)~\cite{Sato:2014uza}, one at  RIKEN-RAL (UK)~\cite{Pizzolotto:2021dai, Pizzolotto:2020fue}, and one at PSI (Switzerland)~\cite{Amaro:2021goz}.
Their goal is to measure with 1~ppm accuracy by means of pulsed laser spectroscopy from which
precise information about the magnetic structure of the proton can be extracted.
Namely, the 2PE  contribution can be extracted with $10^{-4}$ relative accuracy when comparing theory with experiment. 
In a second phase the HFS measurements could reach relative accuracies  below the $10^{-7}$ level, providing very precise benchmark for the understanding of the proton spin structure.
A  measurement of the 1S HFS in $\mu$H, in combination with the precise value for the 1S HFS in H, can be used to disentangle the Zemach radius and proton-polarizability contributions from 2PE, see discussion in Ref.~\cite{Antognini:2022xoo}, and determine them with high precision. This will lead to the best empirical determination of the proton Zemach radius from spectroscopy, without the uncertainty associated with the theoretical prediction of the polarizability contribution, to be compared with determinations from form factor measurements.

For all three collaborations, the main challenge is posed by the small laser-induced transition probability and a transition wavelength at 6.8 $\upmu$m where no adequate (sufficiently powerful) laser technologies are available.
To give a sense of the required technology leap, note that the laser fluence needed for the HFS experiment of the CREMA collaboration is 4 orders of magnitude larger than  used for the measurement of the 2S-2P transition in $\mu$H to determine the proton radius.
Moreover this has to be obtained at a longer wavelength and for much smaller laser bandwidth.
The development of these laser technologies will open the way also for improving the 2S-2P measurements  in $\mu$H and $\mu$D by at least a factor 
 of 5.
This calls for advances on the theory side (BSQED and proton-structure effects).
Similar refined measurements will be possible in  $\mu$He$^+$.



The intrinsic energy resolution of semi-conductor detectors excludes their use for precision measurements on light muonic atoms. However, a 5~ppm precise determination of the $\mu^{12}$C and $\mu^{13}$C 2P-1S transition energies using a crystal spectrometer, and the derived charge radius with an accuracy of $2\cdot 10^{-3}\,$fm, has demonstrated that an x-ray detector with an energy resolution of order 10~eV does enable charge radii measurements to a precision of $10^{-3}$ or better for $Z=3-8$ nuclei. 
Novel cryogenic microcalorimeters (MCs) are a promising candidate to fill this gap between laser spectroscopy and x-ray spectroscopy with solid-state detectors. In recent years MCs have demonstrated energy resolutions of a few to a few tens of eV in the range $1-400\,$keV \cite{2014-6keV,2016-MMCres,2018-MMC5.9, 2020-Th, 2019-384keV}; close to that of crystal spectrometers, while enjoying a higher efficiency and covering a much broader energy range.
With such resolution and the appropriate calibration, line centers can be determined to better than $10\,$ppm \cite{2020-7ppm}.
Achieving this accuracy for measurements of 2P-1S transitions in light muonic atoms, combined with theoretical calculations of nuclear structure and polarizability detailed above, could improve radii determinations by up to an order of magnitude (see. Fig.~\ref{fig:rad}). The same method can also be used for strong-field BSQED tests using Rydberg states in heavier elements.
To pursue these promising opportunities, a new collaboration has formed which will use metallic magnetic calorimeters at PSI.

As for low-lying states at higher $Z(>10)$, solid state detectors are again competitive. A new experimental program at PSI is measuring muonic x-rays from radioactive isotopes using high-purity germanium detectors [see separate contribution ``Muonic atoms for fundamental and applied nuclear science''].

\bibliography{lowQ}
\end{document}